\documentclass[aps,pre,reprint,onecolumn,notitlepage,groupedaddress]{revtex4-1}
\usepackage{epsfig,graphics,amssymb,amsmath,subeqnarray,color}
\usepackage[sfdefault=cmbr]{isomath}

\usepackage{bm}
\def\boldsymbol{\bm}

\def\para{\parallel}

\def \t{\tensorsym}
\def \lb{\left}
\def \rb{\right}

\def \d{\,\text{d}}

\def \betad{\dot{\beta}}

\def \bnabla{\boldsymbol{\nabla}}

\def \bsigma{\boldsymbol{\sigma}}

\def \thetad{\dot{\theta}}

\def \bOmega{\mathbf{\Omega}}

\def \bOmegat{\tilde{\mathbf{\Omega}}}

\def \bTheta{\mathbf{\Theta}}

\def \bXi{\mathbf{\Xi}}

\def \Re{\text{Re}}

\def \bF{\mathbf{F}}

\def \bG{\mathbf{G}}

\def \bI{\mathbf{I}}

\def \bL{\mathbf{L}}

\def \bR{\mathbf{R}}

\def \bT{\mathbf{T}}
\def \bU{\mathbf{U}}
\def \bUt{\tilde{\mathbf{U}}}

\def \bc{\mathbf{c}}

\def \be{\mathbf{e}}

\def \bf{\mathbf{f}}

\def \bk{\mathbf{k}}
\def \bkh{\hat{\mathbf{k}}}

\def \bn{\mathbf{n}}

\def \br{\mathbf{r}}
\def \brt{\tilde{\mathbf{r}}}
\def \brtd{\dot{\tilde{\mathbf{r}}}}

\def \bu{\mathbf{u}}
\def \but{\tilde{\mathbf{u}}}

\def \bx{\mathbf{x}}

\def \by{\mathbf{y}}

\def \St{\tilde{S}}

\def \rd{\dot{r}}

\def \thetad{\dot{\theta}}

\def \bnablat{\tilde{\boldsymbol{\nabla}}}

\def \bzero{\mathbf{0}}

\def \brd{\dot{\br}}
\def \brt{\tilde{\br}}
\def \brtd{\dot{\tilde{\br}}}

\def \but{\tilde{\bu}}

\def \bxd{\dot{\bx}}
\def \bxh{\hat{\bx}}

\def \fL{\mathcal{L}}

\def \fO{\mathcal{O}}

\def \tzero{\mathsf{\t 0}}

\def \tF{\mathsf{\t F}}

\def \tR{\mathsf{\t R}}
\def \tRt{\tilde{\mathsf{\t R}}}

\def \tT{\mathsf{\t T}}
\def \tTt{\tilde{\mathsf{\t T}}}
\def \tU{\mathsf{\t U}}
\def \tUt{\tilde{\mathsf{\t U}}}

\def \tTheta{\mathsf{\t \Theta}}

\begin{document}

\title{A note on the reciprocal theorem for the swimming of simple bodies}
\author{Gwynn J. Elfring\footnote{Corresponding author. Email: gelfring@mech.ubc.ca}}
\affiliation{
Department of Mechanical Engineering, Institute of Applied Mathematics, 
University of British Columbia,
2054-6250 Applied Science Lane, Vancouver, B.C., V6T 1Z4, Canada}

\date{\today}
\begin{abstract}
The use of the reciprocal theorem has been shown to be a powerful tool to obtain the swimming velocity of bodies at low Reynolds number. The use of this method for lower-dimensional swimmers, such as cylinders and sheets, is more problematic because of the undefined or ill-posed resistance problems that arise in the rigid-body translation of these shapes. Here we show that this issue can be simply circumvented and give concise formulas obtained via the reciprocal theorem for the self-propelled motion of deforming two-dimensional bodies. We also discuss the connection between these formulae and Fax\'en's laws.
\end{abstract}

\maketitle

\section{Introduction}
The locomotion of microorganisms in viscous fluids is constrained by the subdominance of inertial forces compared to viscous forces in the flows they generate. Some swimming strategies, such as reciprocal motion (in other words, those which display a time-reversal symmetry) can be effective at large scales while largely fruitless at small scales due to the kinematic reversibility of the Stokes equations \cite{purcell77}. In order to break time-reversal symmetry, microorganisms often pass deformation waves \cite{brennen77}. Determining the swimming kinematics of an organism undergoing a periodic deformation has been a subject of study since Taylor's classic work on the swimming speed of an infinite, two-dimensional sheet passing waves of transverse displacement \cite{taylor51}, and continues to be an active area (see for example two recent reviews \cite{lauga09b,pak15}, and references therein).

Determining the motion of a swimmer in a fluid can be substantially simplified by appealing to the Lorentz reciprocal theorem \cite{happel65}. Stone and Samuel showed that the swimming kinematics may be found without solution of the flow field generated by the swimmer if one has knowledge of the stress on a body of the same instantaneous shape undergoing rigid-body motion only \cite{stone96}. This tool has been subsequently used to simplify a large number of problems from jet propulsion without inertia \cite{spagnolie10b}, to electrokinetic flow enhancement \cite{squires08}, to the motion of surfactant covered droplets in a background flow \cite{pak14}, as a few examples amongst many.

Like Taylor's swimming sheet calculation, many models use simplified geometries or lower dimensional geometries to gain physical understanding or mathematical tractability (for example \cite{elfring09, crowdy13, yazdi14}). The use of the reciprocal theorem for swimming, following Stone and Samuel \cite{stone96}, requires the solution of the stress under rigid-body motion which poses a problem for lower dimensional geometries where rigid-body motion is either ill-posed or undefined. However, unlike the case of rigid-body motion, force-free swimming is well defined even for lower dimensional geometries, and we show here that these problems may be circumvented within the reciprocal theorem framework as well. We also show that by applying the reciprocal theorem to surfaces which are not material we can easily obtain the swimming kinematics for bodies whose shape is given by small deformations from a simple reference surface. Finally we rederive the formulas obtained for swimming via a direct integration of boundary integral formulation of the Stokes equation and discuss the connection with Fax\'en's laws.

\section{Swimmer motion}
\subsection{Boundary motion}
\begin{figure}[b]
\centering
\includegraphics[width=0.65\textwidth]{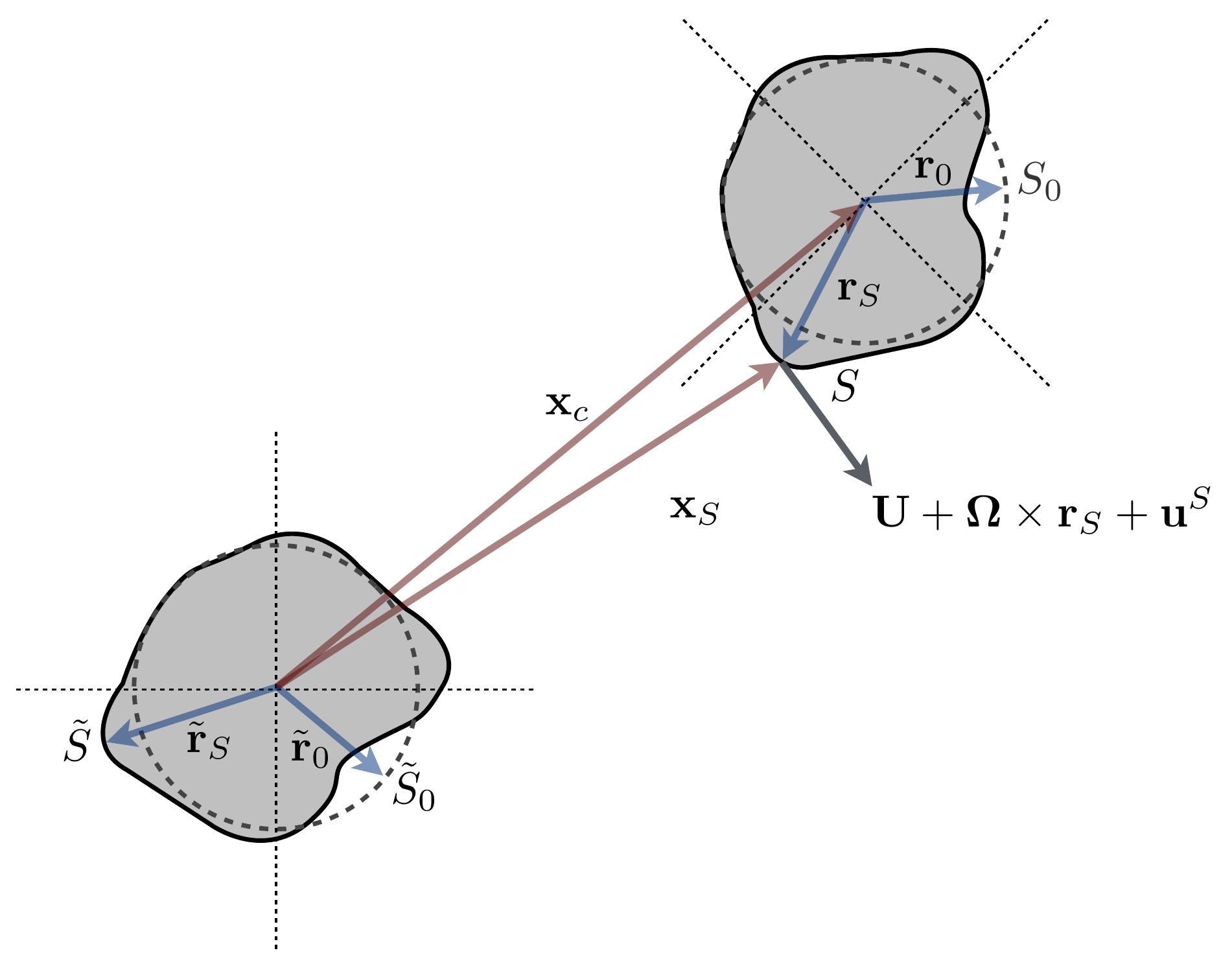}
\caption{Schematic representation of a general swimmer. A position on the surface, $S$, of a swimmer in the lab frame is $\bx_S$ which is found by a translation (by $\bx_c$) and rotation of a point $\brt_S$ on a virtual swimmer. The instantaneous velocity of a point on the surface is given by translation $\bU$, rotation $\bOmega\times\br_S$, and deformation $\bu^S$.}
\label{swimmer}
\end{figure}

In order to swim, a body undergoes a periodic deformation of its surface $S(t)$. To describe the swimming gait of the body we use a virtual \cite{ishimoto12} (or unlocated \cite{shapere89}) body which does not move in response to the fluid. A point on the surface of the virtual swimmer, $\St(t)$, is given by $\brt_S(t)$ and the deformation of this surface in time is referred to as the swimming gait. Periodic deformations may be described as deviations from a reference surface
\begin{align}
\brt_S(\br_0,t)- \brt_0 =\Delta\brt(\brt_0,t),
\end{align}
where $\brt_0$ is a point on the reference surface, $\St_0$, of the unlocated body. When such a deforming body is submerged in a fluid, stresses exerted by the fluid lead generally to a rigid-body translation and rotation of the body. The position of a point on the surface of the (located) real swimmer, $\bx_S\in S(t)$, is hence given by the translation, by $\bx_c(t)$ and rotation by $\bTheta(t)$ of that position on the virtual body, $\brt_S$ (see Fig.~\ref{swimmer}), hence 
\begin{align}
\bx_S(\brt_0,t) = \bx_c(t)+\bTheta\cdot\brt_S(\brt_0,t).
\end{align}
A point on the reference surface in the lab frame, $\bx_0 \in S_0$, is similarly writen $\bx_0=\bx_c+\bTheta\cdot\brt_0$. For simplicity we introduce the notation $\br_{S}=\bTheta\cdot\brt_S$ and $\br_0=\bTheta\cdot\brt_0$.

Upon differentiation of the position of a point on the body we obtain the velocity
\begin{align}\label{swimmerbc}
\bu(\bx_S)=\frac{\partial\bx_S}{\partial t}=\bU+\bOmega\times\br_S+\bu^S.
\end{align}
We have defined $\bU = \frac{\d\bxd_c}{\d t}$ and $\bu^S=\bTheta\cdot\frac{\partial\brt_S}{\partial t}$, while the rotation operator is related to the angular velocity, $\bOmega$, by the relationship  $\frac{\d \bTheta}{\d t} =\bOmega\times\bTheta$. The first two terms represent rigid-body motion while the third term represents the deformation of the surface due to the swimming gait \cite{felderhof12,elfring15}. We note that the instantaneous translation and rotation of the body depends on the choice of reference point on the body, a satisfying resolution of this ambiguity is to specify that $\bx_c$ be the center-of-mass of the body and then require that the gait of the virtual swimmer conserve momentum \cite{ishimoto12}, but this complication is often unnecessary, in particular if one is ultimately interested in simply the time-averaged swimming speed. An alternative and equivalent description is to use a body-fixed basis to describe the deformation of the swimmer \cite{yariv06}.

We describe here the locomotion of microorganisms small enough such that the Reynolds number of the flows generated may be taken to be zero and hence the Stokes equations are applicable
\begin{align}
\bnabla\cdot\bsigma=-\bnabla p +\eta\nabla^2\bu = \bzero, \quad\quad \bnabla\cdot\bu=0,
\end{align}
and the bodies are instantaneously force and torque free,
\begin{align}
\bF = \int_S \bn\cdot\bsigma \d S &=\bzero,\\
\bL = \int_S \br_S\times(\bn\cdot\bsigma) \d S &=\bzero,
\end{align}
where the surface $S$ is a function of time and the normal to the surface $\bn$ points into the fluid while $\bsigma$ is the stress tensor. For compactness we use six dimensional vectors which contain both force and torque $\tF = [\bF \ \ \bL]^\top$ and translation and rotation $\tU = [\bU \ \ \bOmega]^\top$.

Due to the linearity of the Stokes equations for a Newtonian fluid, the flow field arising from a force- and torque-free body undergoing a surface deformation may be decomposed into two parts $\bu=\bu_D+\bu_T$. One part in which the body is undergoing only rigid-body motion $\bu_D(\bx_S) = \bU+\bOmega\times\br_S$, and another part in which the boundary motion is due to the swimming gait alone, $\bu_T(\bx_S) = \bu^S$ \cite{elfring15}. There is, generally, a net force and torque in each problem. Under rigid-body motion, the drag $\tF_D=-\tR\cdot\tU$, where $\tR$ is the rigid-body resistance tensor, while if the deforming body is held fixed ($\tU=\tzero$), it generates an instantaneous `thrust' (or swim force \cite{takatori14}), $\tF_T$. The sum of these forces must be zero, $\tF_D+\tF_T=\tzero$ and so we write
\begin{align}\label{prob1}
\tU = \tR^{-1}\cdot\tF_T.
\end{align}
In order to utilize this formula one must solve both the rigid-body motion problem, $\bu_D$ (to obtain $\tR$), and the thrust problem, $\bu_T$ (to obtain $\tF_T$). However, in a landmark paper \cite{stone96}, Stone and Samuel showed that by appealing to the reciprocal theorem \cite{happel65}, the solution of the flow field $\bu_T$ may be avoided entirely.

\section{The reciprocal theorem}\index{Reciprocal theorem}
We denote $\bsigma_D$ as the stress tensor associated with the velocity field due to rigid-body motion, $\bu_D$, while $\bsigma_T$ as the stress tensor for associated with the velocity field due to deformation, $\bu_T$.  The stress and velocity fields satisfy the relationship $\bnabla\cdot\bsigma_D\cdot\bu_T=\bnabla\cdot\bsigma_T\cdot\bu_D=0$ and by integrating over the fluid volume exterior to $S$ by means of the divergence theorem \cite{stone96}, one finds
\begin{align}\label{rt1}
\tF_T\cdot\tU&=\int_S \bn\cdot\bsigma_D\cdot\bu^S\d S.
\end{align}
Due to the linearity of the Stokes equations we may write $\bsigma_D = \tT\cdot\tU$ and hence the resistance tensor is given by the relationship
\begin{align}
\tR =-\int_S
\begin{bmatrix}
\bn\cdot\tT\\
\br_S\times\lb(\bn\cdot\tT\rb)
\end{bmatrix}
\d S.
\end{align}
Substituting into \eqref{rt1} we find
\begin{align}
\tF_T &=\int_S \bu^S\cdot(\bn\cdot\tT)\d S,
\end{align}
indicating that the hydrodynamic `thrust' $\tF_T$ may be written as an integral of the rigid-body stress tensor weighted by the gait $\bu^S$. Finally substitution into \eqref{prob1} gives the rigid-body motion of the swimmer
\begin{align}\label{locomotion}
\tU
= \tR^{-1}\cdot\lb[\int_S\bu^S\cdot(\bn\cdot\tT)\d S\rb],
\end{align}
as shown by Stone and Samuel \cite{stone96}. In this form we see that the swimming velocity is a simple linear functional of the gait, $\tU =\fL(\bu^S)$, from which well known constraints on swimming in Stokes flows such as the scallop theorem or the rate invariance of the distance travelled in a stroke follow directly \cite{ishimoto12}. 
Note that for two-dimensional bodies in Stokes flows the resistance, or dragging, problem may be ill-posed; the stress on the surface of a translating cylinder, for example, is singular in the Reynolds number, nevertheless the product $\bn\cdot\tT\cdot\tR^{-1}$ is well defined and hence this formalism may be used for lower dimensional bodies. Finally we may recast the swimming motion with respect to the unoriented body
\begin{align}\label{locomotionun}
\tU=\tTheta\cdot\tUt
= \tTheta\cdot\lb\{\tRt^{-1}\cdot\lb[ \int_{\St} \frac{\partial \brt_S}{\partial t}\cdot(\bn\cdot\tTt)\d S\rb]\rb\},
\end{align}
where $\tUt=[\bUt \ \ \bOmegat]^\top$ indicates unoriented rigid-body motion and the six dimensional rotation operator,
\begin{align}
\tTheta=
\begin{bmatrix}
\bTheta && \bzero\\
\bzero && \bTheta
\end{bmatrix}.
\end{align}

\section{Simple bodies}
Analytical progress can be made by considering shapes which align with a coordinate system. We describe below the swimming of bodies with spherical, cylindrical and planar shape. The spherical case was demonstrated by Stone and Samuel \cite{stone96}. The cylindrical and spherical cases show that the undefined or singular difficulties with the rigid-body resistance problem can be circumvented to yield formulas equivalent to the spherical case.

\subsection{Spherical swimmers}
If the body is a sphere of radius $a$, then the rigid-body problem is well known; the resistance matrix is diagonal and hence easily invertible while $\bn\cdot\tT = -\frac{3\eta}{2a}[\bI \ \ 2\bXi]$ where $\bXi=-\br\times\bI$ where $\bI$ is the identity tensor. With these formulas the swimming velocity for a sphere with a prescribed velocity field, $\bu^S$, on its surface is
\begin{align}
\tU = -\frac{1}{4\pi a^2}\int_S
\begin{bmatrix}
\bI\\
\frac{3}{2a^2}\bXi^\top
\end{bmatrix}\cdot\bu^S\d S
= -\begin{bmatrix}
\lb<\bu^S\rb>\\
\frac{3}{2a^2}\lb<\br_S\times\bu^S\rb>
\end{bmatrix},
\end{align}
(where the angle brackets denote an average over the surface $S$) as shown by Stone and Samuel \cite{stone96}.

\subsection{Cylindrical swimmers}

Consider the flow around a rigidly translating and rotating cylinder of radius $a$. The cylinder can translate in any direction but we restrict rotations to the axial direction ($\bOmega_\para$). The surface traction may be written as
\begin{align}
\bn\cdot\tT = -\frac{\eta}{a}\lb[\beta\bI_\para+\alpha\bI_\perp \ \ 2\bXi\cdot\bI_\para\rb]
\end{align}
We use the $\parallel$ subscript to denote the axial direction while $\perp$ denotes components perpendicular. For translation normal to the surface of the cylinder, $\alpha$ is a dimensionless constant which is singular in $\Re$ \cite{pozrikidis11}. For axial translation we may say we have an outer cylinder of radius $R$ at infinity and so $\beta = \lim_{R\rightarrow\infty}\frac{a}{R-a}$. Integrating over the surface of the cylinder we obtain the resistances per unit length
\begin{align}
\bR_{FU}^{-1} &= \frac{1}{2\pi\eta}\lb[\beta^{-1}\bI_\para+\alpha^{-1}\bI_\perp\rb],\\
\bR_{L\Omega}^{-1} &= \frac{1}{4\pi\eta a^2}\bI_\perp.
\end{align}
Combining these terms we obtain the swimming velocity
\begin{align}
\tU = -\begin{bmatrix}
\lb<\bu^S\rb>\\
\frac{1}{a^2}\lb<\br_S\times\bu^S\rb>\cdot\bI_\para
\end{bmatrix}.
\end{align}
Notice that the result is independent of $\alpha$ and $\beta$ and thus the singular nature of the resistance problem is avoided because the product $\bn\cdot\tT\cdot\tR^{-1}$ is well defined. Squires and Bazant first derived the form of the solution for the translation of an infinite cylinder perpendicular to its axis, in the context of electrophoretic migration, by noting that $\bn\cdot\bsigma_D$ is constant \cite{squires06}.

\subsection{Planar swimmers}
Consider a flat two dimensional sheet which may have a different prescribed velocity on each side of the sheet, unequally spaced between two rigid walls. If the sheet can only move in its own plane $\bU_\para$, the resistance problem would be shear flow where
\begin{align}
\bn\cdot\bsigma_D = -\frac{\eta}{h}\bU_\para.
\end{align}
If the distances between the two walls are $h_1$ and $h_2$ we have then as the only non-zero mobility
\begin{align}
\bR_{FU}^{-1} = \frac{1}{\eta A}\lb[h_1^{-1}+h_2^{-1}\rb]^{-1}\bI_\para,
\end{align}
and with this Eq.~\eqref{locomotion} becomes
\begin{align}\label{twowalls}
\tU = \frac{-1}{h_1+h_2}\begin{bmatrix}
\bI_\parallel\\
\bzero
\end{bmatrix}
\cdot\bigg(h_2\lb <\bu^{S_1}\rb>
+h_1\lb<\bu^{S_2}\rb>
\bigg).
\end{align}
In an unbounded fluid, $h_1=h_2\rightarrow \infty$, we obtain
\begin{align}\label{sheetvelocity}
\tU = -\frac{1}{2}\begin{bmatrix}
\bI_\parallel\\
\bzero
\end{bmatrix}
\cdot\bigg(\lb <\bu^{S_1}\rb>
+\lb<\bu^{S_2}\rb>
\bigg) = -
\begin{bmatrix}
\bI_\parallel\\
\bzero
\end{bmatrix}
\cdot\lb<\bu^S\rb>.
\end{align}
If the the sheet is symmetric, $\lb<\bu^{S_1}\rb>=\lb<\bu^{S_2}\rb>$ and one need only solve the half space problem.

\subsection{General form}
All the formulas derived above for finding the swimming velocity of an unbounded sphere, cylinder or sheet, $\tU$ due to a prescribed boundary condition, $\bu^S$, take the general form
\begin{align}\label{surfaceaverage}
\tU = -\begin{bmatrix}
\lb<\bu^S\rb>\\
C\lb<\br_S\times\bu^S\rb>
\end{bmatrix},
\end{align}
where $C$ is simply a geometric factor. This may equivalently be written as
\begin{align}
\lb<\bu\rb>&=\bzero,\label{averageu}\\
\lb<\br_S\times\bu\rb>&=\bzero,\label{averageru}
\end{align}
where we see that because of the symmetry of the stress, both the surface-averaged velocity and the surface-averaged first moment of the velocity must be zero. This connection was demonstrated for a sphere by Felderhof \cite{felderhof12}. Equations \eqref{averageu} and \eqref{averageru} may be obtained through a direct integration of the boundary integral formulation of the Stokes equations over the surface of a sphere, or a cylinder following Batchelor's approach to the derivation of Fax\'en first law (see Appendix) \cite{batchelor72}. For a flat sheet, we observe that the force on the surface is given simply by
\begin{align}
\bF = \eta\lb<\frac{\partial\bu}{\partial n}\rb>=\bzero,
\end{align}
which is indeed true on every plane parallel to the sheet and so we must have
\begin{align}
\lb<\bu\rb> =\bzero,
\end{align}
on the surface of the sheet. The constant of integration is determined by the far-field condition.

Recall that $\bu(\bx_S) = \bU +\bOmega\times\br_S + \bu^S$ and so upon substitution into \eqref{averageu} and \eqref{averageru} we find
\begin{align}
\bU &= -\lb<\bu^S\rb>,\\
a^2\lb<\bI-\bn\bn\rb>\cdot\bOmega &= -\lb<\br_S\times\bu^S\rb>.
\end{align}
For a sphere $\lb<\bI-\bn\bn\rb>=\frac{2}{3}\bI$ and so $C=3/2a^2$ in Eq.~\eqref{surfaceaverage}, while for a cylinder where $\bOmega=\bOmega_\para$, hence $\lb<\bI-\bn\bn\rb>\cdot\bOmega=\bOmega$ and so $C=1/a^2$.

\subsection{Deforming bodies}
We have described above how to obtain the motion for objects with simple shapes which align with coordinate axes. If the time dependent surface of a swimmer, $S(t)$, deviates only slightly from a simple reference surface, $S_0$, then we can, through Taylor series recast the problem to obtain the velocity field on $S_0$ \cite{felderhof94a,felderhof94b}. The velocity field is expanded
\begin{align}\label{velonbody}
\bu(\bx_S)&=\bu(\bx_0)+(\bx_S-\bx_0)\cdot\lb.\bnabla\bu\rb|_{\bx_0}+\fO(|\Delta\br|^2),
\end{align}
then rearranging we obtain the boundary condition on $S_0$,
\begin{align}\label{bcsmall}
\bu(\bx_0) = \bU +\bOmega\times\br_0+\bu^{S_0},
\end{align}
where
\begin{align}\label{speedS0}
\bu^{S_0}&=\frac{\partial\Delta\br}{\partial t}+ \bOmega\times\Delta\br-\Delta\br\cdot\lb.\bnabla\bu\rb|_{\bx_0}+\fO(|\Delta\br|^2),
\end{align}
and $\Delta\br = \br_S-\br_0$. We remove the dependence on orientation of this instantaneous boundary condition by taking the product of both sides with $\bTheta^\top$ to obtain
\begin{align}
\but(\br_0)=\but^{\St_0} = \frac{\partial\Delta\brt}{\partial t}+ \bOmegat\times\Delta\brt-\Delta\brt\cdot\lb.\bnablat\but\rb|_{\brt_0}+\fO(|\Delta\brt|^2),
\end{align}
where $\bOmegat = \bTheta^\top\cdot\bOmega$ and $\but(\brt)=\bTheta^\top\cdot\bu(\bx)$ is the velocity field with respect to the unlocated body with $\bx=\bx_c+\bTheta\cdot\brt$ .

This formulation leads to a swimming problem defined entirely on $S_0$ which satisfies the correct boundary conditions on $S$. In order to solve for the swimming kinematics, we apply the reciprocal theorem on $S_0$, namely we decompose the problem into finding the resistance to the rigid-body translation and rotation of $S_0$ and the thrust force generated by $\bu^{S_0}$. Nothing in the formulation of the reciprocal theorem for swimming prevents its application on a surface that is not material and by doing so one avoids the expansion of a surface integral over a non-trival domain \cite{lauga14} and in addition now we need only to solve for a single resistance problem. For bodies that are nearly spherical, cylindrical or planar in particular we may write
\begin{align}
\tU = -\begin{bmatrix}
\lb<\bu^{S_0}\rb>\\
C\lb<\br_0\times\bu^{S_0}\rb>
\end{bmatrix}
= -
\tTheta
\cdot
\begin{bmatrix}
\lb<\but^{\St_0}\rb>\\
C\lb<\brt_0\times\but^{\St_0}\rb>
\end{bmatrix}=\tTheta\cdot\tUt.
\label{velrel}
\end{align}
One may first determine $\tUt$ and then integrate $\bOmegat$ to obtain the orientation in the lab frame. Note that the swimming gait on $\St_0$, $\but^{S_0}$, depends on gradients of the (unknown) flow field $\but$ and the rotation rate of the swimmer $\bOmegat$. However if we take $\Delta \brt = \sum_{m>0} \epsilon^m\brt_m$ where $\epsilon \ll 1$ is a dimensionless measure of gait amplitude and then consequently expand the velocity field, $\but=\sum_m \epsilon^m\but_m$, we overcome these difficulties perturbatively, at each order
\begin{align}
\but_1^{\St_0}&=\frac{\partial\brt_1}{\partial t},\label{bcformula1}\\
\but_2^{\St_0}&=\frac{\partial\brt_2}{\partial t}+\bOmegat_1\times\brt_1-\brt_1\cdot\lb.\bnablat\but_1\rb|_{\brt_0}.\label{bcformula2}
\end{align}
This yields the instantaneous boundary condition for the velocity field on $\St_0$ in the orientation order by order and then by Eq.~\eqref{velrel} we obtain $\tUt$.

\section{Example Model Swimmers}
In the following section we revisit two classic model swimmers, Taylor's swimming sheet \cite{taylor51} and a squirming \cite{lighthill52} cylinder \cite{blake71c}, using the reciprocal theorem formalism presented above.

\subsection{Deforming sheet}
We describe here an infinite two-dimensional sheet with time varying deformation of its surface from a planar reference surface. Due to its infinite extent the body does not rotate, $\bTheta(t)=\bTheta(0)$, and we take $\bTheta(0)=\bI$ for convenience. If the deformation is in the form of periodic waves then we may write
\begin{align}
\Delta\br= A\sum_{n} \bc_n e^{inz}.
\end{align}
The vector of complex Fourier coefficients, $\bc_n$, delineates wave shape ($\bc_0=\bzero$), while the wave variable, $z = \bk\cdot\br_0 - \omega t$, depends on the wave vector, $\bk$, and the actuation frequency, $\omega$.  The boundary conditions for an extensible sinusoidal sheet, $\bc_1 = -i/2\be_y$, and an inextensible sheet were first described by Taylor \cite{taylor51}, and catalogued in detail elsewhere \cite{elfring11}.

Non-dimensionalizing lengths by $|\bk|$ and time by $\omega$ we obtain
\begin{align}
\Delta\br'= \epsilon\sum_n \bc_n e^{inz}=\epsilon\br'_1,
\end{align}
where $\epsilon = A|\bk|$ and primes indicate dimensionless quantities. Dropping the primes and assuming dimensionless variables we obtain the boundary conditions on $S_0$ order by order in $\epsilon$, 
\begin{align}
\bu_1^{S_0} &= -\sum_nni  \bc_n e^{niz},\\
\bu_2^{S_0} &= -\sum_n  e^{inz}\bc_n\cdot\lb.\bnabla\bu_1\rb|_{\bx_0}.
\end{align}
We note that for this general deformation, the values of $\bu^{S_0}$ must be computed for each side of the sheet $S_1$ and $S_2$ as the value of the mean need not be equal.

Expressing the swimming velocity $\bU(\epsilon) = \sum \epsilon^m\bU_m$ and using Eq.~\eqref{sheetvelocity} for an unbounded swimmer we see immediately that due to the wave-like deformation
\begin{align}
\bU_1 = -\lb<\bu_1^{S_0}\rb>=\bzero.
\end{align}
To find the swimming speed at second order we must obtain the complete first order flow field in order to evaluate the gradients $\bnabla\bu_1$. The exception is if the sheet is only deforming tangentially in the plane of $S_0$, in that case only in-plane derivatives, $\bnabla_0$, are required
\begin{align}
\bnabla_0\bu_1 = \bnabla_0\brd_1=\sum_n n^2 \bkh\bc_ne^{inz}.
\end{align}
The reciprocal theorem for an unbounded swimmer, Eq.~\eqref{sheetvelocity}, then yields
\begin{align}\label{tangential}
\bU_2=-\lb<\bu_2^{S_0}\rb>=\sum_n n^2\bc_n^\dagger\bc_n\cdot\bkh
\end{align}
where $^\dagger$ denotes the complex conjugate. Interesting to note is that when the sheet is subject to tangential deformations only, the presence of walls is strictly irrelevant, provided the prescribed velocity field is the same on both sides of the sheet, as described in Eq.~\eqref{twowalls}.

To include out-of-plane deformations of the sheet we must solve for $\bu_1$. At this point we'll assume only fully two-dimensional flow fields with $\bc_n = a_n\be_x+b_n\be_y$ where $\be_x = \bk/|\bk|$ while $\be_y$ is normal to the sheet. The flow field may be obtained by means of the streamfunction $\bu=\bnabla^\perp\psi$ and the general case follows directly from classical examples for single modes as each Fourier mode is decoupled at leading order (details are hence omitted). The mean of the boundary condition on each side of the sheet is given by
\begin{align}
\lb<\bu_2^{S_{1,2}}\rb>&=-\sum_n n^2\lb[a_na_n^\dagger\pm A(nh)ia_nb_n^\dagger-B(nh) b_nb_n^\dagger\rb]\be_x,\label{sheetbc2}
\end{align}
where the sign corresponds to the surface with normal $\bn = \pm\be_y$, and
\begin{align}
A(x) = \frac{\sinh(2x)-2x}{\sinh^2 (x)-x^2},\quad B(x) = \frac{\sinh^2 (x)+x^2}{\sinh^2 (x)-x^2}.
\end{align}
This difference in the mean value of the boundary condition on each side of an asymmetric sheet was recently noted by Felderhof \cite{felderhof14}. Using Eq.~\eqref{sheetbc2} we obtain the swimming speed of a sheet that is unevenly spaced between two walls directly from Eq.~\eqref{twowalls}
\begin{align}
\bU_2 = \sum_nn^2\Bigg[a_na_n^\dagger &+\frac{h_2A(nh_1)-h_1A(nh_2)}{h_1+h_2}ia_nb_n^\dagger
-\frac{h_2B(nh_1)+h_1B(nh_2)}{h_1+h_2}b_nb_n^\dagger\Bigg]\be_x.
\end{align}
Without out-of-plane deformations, ($b_n=0$), this result matches Eq.~\eqref{tangential} and indeed the motion of the sheet is unaffected by the presence of a wall. It is important to note here that this result is inconsistent with that found by Katz \cite{katz74} for a single mode swimmer near two walls due to the minus sign in the second term which arises from the asymmetric boundary condition noted in Eq.~\eqref{sheetbc2}. Now with only out-of-plane deformations, $a_n=0$, the sheet is symmetric and we obtain Reynolds result \cite{reynolds65} for a single non-zero mode. In the particular case of an unbounded swimmer, $h\rightarrow \infty$, while $A(nh)\rightarrow 2\text{sgn}(n)$ and $B(nh) \rightarrow 1$ and so the swimming velocity is given by
\begin{align}
\bU_2=\sum_n n^2\lb[a_n a_n^\dagger-b_n b_n^\dagger \rb]\be_x.
\end{align}
This differs with the general result first found by Blake \cite{blake71b}, again due to the asymmetry in the boundary condition on either side of the sheet, specifically because the term $\pm2i\text{sgn}(n)a_nb_n^\dagger$ ultimately cancels out when we average the top and bottom of the sheet. Felderhof argues that for a finite circular disk the velocity difference across the disk induces a rotation of the body \cite{felderhof14}. Taylor's sinusoidal swimming sheet, where the only non-zero mode is $b_1=-i/2$, yields the classic result $\bU = -\frac{1}{2}\epsilon^2\be_x$ to leading order \cite{taylor51}.

\subsection{Deforming cylinder}
Following Blake \cite{blake71c}, we describe the deformation of the surface of a cylinder through material coordinates on the virtual swimmer $r_S(\theta_0,t)=r_0 + \epsilon r_1(\theta_0,t)$ and $\theta_S(\theta_0,t) = \theta_0+\epsilon\theta_1(\theta_0,t)$, then a position on the material surface in this frame is simply $\brt_S = r_S\be_r(\theta_S)$. By expanding in $\epsilon$ we obtain
\begin{align}
\Delta\brt =\sum_{m>0}\epsilon^m\brt_m &=\epsilon (r_1\be_r+r_0\theta_1\be_\theta)+\epsilon^2(r_1\theta_1\be_\theta-(1/2)r_0\theta_1^2\be_r)+...
\end{align}
where the convention is $\be_r \equiv \be_r(\theta_0)$. Upon differentiation we have $\brtd_1 = \rd_1\be_r+r_0\thetad_1\be_\theta$ and $\brtd_2 = (\rd_1\theta_1+r_1\thetad_1)\be_\theta-r_0\theta_1\thetad_1\be_r$. The motion of the body then follows from the equations \eqref{bcformula1} and \eqref{bcformula2}, to leading order we have 
\begin{align}
\tUt_1 &= -\begin{bmatrix}
\lb<\brtd_1\rb>\\
r_0^{-2}\lb<\brt_0\times\brtd_1\rb>
\end{bmatrix}
=-\begin{bmatrix}
\lb<\rd_1\be_r\rb>+r_0\lb<\thetad_1\be_\theta\rb>\\
\lb<\thetad_1\rb>\be_z
\end{bmatrix}.
\end{align}
A proper swimming gait is periodic in time and so a time-averaged velocity in the swimming frame is zero, $\overline{\tUt_1} = \tzero$. Of course the time-averaged velocity of the swimmer need not be zero in the lab frame over the same period as rotation and translation are coupled. The leading-order rotation vanishes however, if the tangential deformation has zero mean, as we now assume.

To obtain locomotion at quadratic order one needs to solve for the entire velocity field at leading order unless only tangential deformation occurs, in other words $r_1=0$, yielding a so-called squirmer. In this case we obtain 
\begin{align}
\tUt_2 &= -\begin{bmatrix}
\lb<\brtd_2\rb>-\lb<\brt_1\cdot\bnablat_0\brtd_1\rb>\\
r_0^{-2}\lb[\brt_0\times\brtd_2-\lb<\brt_0\times(\brt_1\cdot\bnablat_0\brtd_1)\rb>\rb]
\end{bmatrix}
=\begin{bmatrix}
r_0\lb<\theta_1\partial_{\theta_0}\thetad_1\be_\theta\rb>\\
\lb<\theta_1\partial_{\theta_0}\thetad_1\rb>\be_z
\end{bmatrix}.
\end{align} 
Exploiting the periodicity of the geometry one can write the deformation $\theta_1 = \sum_n b_n(t) e^{in\theta_0}$. We see then that
\begin{align}
\tUt_1 &=-\begin{bmatrix}
r_0\Im[\dot{b}_1(t)]\be_x+r_0\Re[\dot{b}_1(t)]\be_y\\
\bzero
\end{bmatrix},
\end{align}
while
\begin{align}
\tUt_2 &=-\begin{bmatrix}
\frac{r_0}{2}\sum_nn\dot{b}_n\lb[(b_{n+1}^\dagger-b_{n-1}^\dagger)]\be_x+(b_{n+1}^\dagger+b_{n-1}^\dagger)\be_y\rb]\\
\sum_n in\dot{b}_nb_n^\dagger\be_z
\end{bmatrix},
\end{align}
For directed motion it is typical to restrict $\theta_1$ to be an odd function, doing so by taking $b_n = -(i/2)\beta_n$ where the coefficients $\beta_n=-\beta_{-n}$ are real, one obtains up to quadratic order
\begin{align}
\tUt &=\begin{bmatrix}
\epsilon\frac{r_0\betad_1}{2}\be_x+\epsilon^2\frac{r_0}{4}\sum_{n>0}n\betad_n(\beta_{n-1}-\beta_{n+1})\be_x\\
\bzero
\end{bmatrix},
\end{align}
in agreement with Blake's result \cite{blake71c}. Finally, the leading order time-averaged rigid-body motion of the swimmer in the lab frame is then given by
\begin{align}
\overline{\tU} &= \tTheta(0)\cdot\overline{\tUt}= \tTheta(0)\cdot\begin{bmatrix}
\epsilon^2\frac{r_0}{4}\sum_{n>0}n\overline{\betad_n(\beta_{n-1}-\beta_{n+1})}\be_x\\
\bzero
\end{bmatrix}.
\end{align}

\section{Conclusion}
The reciprocal theorem for swimming allows the solution of the kinematics of a swimming microorganism without the resolution of the entire flow-field. Here we've shown the reciprocal theorem can also be applied to study the motion of two-dimensional swimmers. The reciprocal theorem shows that both the surface-averaged velocity and the surface-averaged first moment of the velocity must be zero on swimming sphere, cylinder and sheet, a result that may also be obtained directly by integrating the boundary integral formulation of the Stokes equations.

\acknowledgements
The author thanks Eric Lauga and On Shun Pak for many fruitful discussions and acknowledges funding from Natural Science and Engineering Research Council of Canada (NSERC). The author also wishes to acknowledge the helpful suggestions of the anonymous referees.

%


\appendix
%
%

\section{Boundary integral formulation (sphere)}
The reciprocal theorem was used to obtain the formulas
\begin{align}
\lb<\bu\rb>=\bzero, \quad\quad \lb<\br_S\times\bu\rb>&=\bzero\nonumber.
\end{align}
These formulas for a sphere and a cylinder may be alternatively found by using the boundary integral formulation of the Stokes equations following Batchelor's derivation of Fax\'en's first law \cite{batchelor72}. We also include an arbitrary background flow for full generality.
\subsection{3D boundary integral formulation}
With the boundary integral formulation of the Stokes equations for a point on the boundary $\bx_S$
\begin{align}
\bu(\bx_S) &= \bu_\infty(\bx_S)-\frac{1}{4\pi\eta}\int_{S}\Big[ \bG(\bx_S,\by)\cdot\bf(\by)
+\eta \bu(\by)\cdot\bT(\bx_S,\by)\cdot\bn(\by)\Big]\d S(\by)\nonumber\\
&= \bU+\bOmega\times\br_S+\bu^S,
\end{align}
where $\bu_\infty$ is an arbitrary background flow, the Stokeslet, $\bG$, and its associated stress tensor, $\bT$, are given
\begin{align}
\bG(\bx,\by) = \frac{\bI}{|\bxh|}+\frac{\bxh\bxh}{|\bxh|^3}, \quad\quad \bT(\bx,\by) = -6\frac{\bxh\bxh\bxh}{|\bxh|^5},
\end{align}
where $\bxh=\bx-\by$.

If we now average this velocity field over the surface $S$, written as $\lb<\bu\rb>$, we obtain
\begin{align}
\lb<\bu\rb>= \lb<\bu_\infty\rb>-\frac{1}{4\pi\eta}\int_{S}\Big[ \lb<\bG\rb>\cdot\bf(\by)
+\eta \bu(\by)\cdot\lb<\bT\rb>\cdot\bn(\by)\Big]\d S(\by)=\bU+\lb<\bu^S\rb>. 
\end{align}
Using two identities \cite{batchelor72, pozrikidis92}, for averages over the surface of a sphere,
\begin{align}
\lb<\bG\rb> &= \frac{4}{3a}\bI,\\
\lb<\bT\rb>\cdot\bn(\by) &= \frac{1}{a^2}\bI,
\end{align}
we obtain
\begin{align}
\bU=\lb<\bu_\infty-\bu^S\rb>-\frac{1}{6\pi \eta a}\bF.
\end{align}
If the swimmer is force free, $\bF=\bzero$, moving in an otherwise quiescent fluid, $\bu_\infty=\bzero$, then we have simply
\begin{align}
\lb<\bu\rb>=\bzero.
\end{align}

Similarly
\begin{align}
\lb<\br_S\times\bu\rb>&= \lb<\br_S\times\bu_\infty\rb>-\frac{a}{4\pi\eta}\int_{S}\Big[\lb<\bn\times\bG\rb>\cdot\bf(\by)+\eta\bu(\by)\cdot\lb<\bn\times\bT\rb>\cdot\bn(\by)\Big]\d S(\by),\nonumber\\
&=\lb<(\br_S\cdot\br_S)\bI-\br_S\br_S\rb>\cdot\bOmega+\lb<\br_S\times\bu^S\rb>.
\end{align}
Noting that $\br_S=a\bn$ and utilizing the following identities
\begin{align}
\lb<\bn\times\bG\rb> &= \frac{2}{3a}\bn(\by)\times\bI,\\
\lb<\bn\times\bT\rb>\cdot\bn(\by) &= \frac{1}{a^2}\bn(\by)\times\bI,
\end{align}
leads to
\begin{align}
\bOmega =\frac{3}{2a^2}\lb<\br_S\times\lb(\bu_\infty-\bu^S\rb)\rb>-\frac{1}{8\pi\eta a^3}\bL=\bzero.
\end{align}
For a torque-free swimmer in a quiescent background flow we have simply
\begin{align}
\lb<\br_S\times\bu\rb>=\bzero.
\end{align}

The formulas above show that the rigid-body motion of a spherical swimmer is identical to that of an inert sphere placed into a background flow given by $\bu_\infty-\bu^S$ according to Fax\'en's laws.

\subsection{2D boundary integral formulation (cylinder)}
With the boundary integral formulation of the Stokes equations for a point on the boundary $\bx_S$, where $S$ now refers to the one dimensional perimeter of the cylinder, we have
\begin{align}
\bu(\bx_S)&= \bu_\infty-\frac{1}{2\pi\eta}\int_{S}\Big[ \bG(\bx_S,\by)\cdot\bf(\by)
+\eta\bu(\by)\cdot\bT(\bx_S,\by)\cdot\bn(\by)\Big]\d S(\by)\nonumber\\
&=\bU+\bOmega\times\br_S+\bu^S,
\end{align}
where the 2D Stokeslet $\bG$ and its associated stress tensor $\bT$ are given
\begin{align}
\bG(\bx,\by) = -\bI\ln(|\bxh|)+\frac{\bxh\bxh}{|\bxh|^2}, \quad\quad \bT(\bx,\by) = -4\frac{\bxh\bxh\bxh}{|\bxh|^4}.
\end{align}

Averaging the velocity field over the surface of the cylinder using the identities
\begin{align}
\lb<\bG\rb> &= \lb(\frac{1}{2}-\ln a\rb)\bI,\\
\lb<\bT\rb>\cdot\bn(\by) &= \frac{1}{a}\bI,
\end{align}
we obtain
\begin{align}
\bU=\lb<\bu_\infty-\bu^S\rb>-\frac{\lb(\frac{1}{2}-\ln a\rb)}{4\pi \eta}\bF.
\end{align}
where $\bF$ is a force per unit length. For a force free swimmer in an otherwise quiescent flow we have simply
\begin{align}
\lb<\bu\rb> = \bzero.
\end{align}

Conversely taking the average $\lb<\br_S\times\bu\rb>$ with the identities
\begin{align}
\lb<\bn\times\bG\rb> &= \bn(\by)\times\bI,\\
\lb<\bn\times\bT\rb>\cdot\bn(\by) &= \frac{1}{a}\bn(\by)\times\bI,
\end{align}
leads to
\begin{align}
\bOmega = \frac{1}{a^2}\lb<\br_S\times\lb(\bu_\infty-\bu^S\rb)\rb>-\frac{1}{4\pi\eta a^2}\bL,
\end{align}
where $\bL$ is torque per unit length. For a torque free swimmer in an otherwise quiescent field we have simply
\begin{align}
\lb<\br_S\times\bu\rb>=\bzero.
\end{align}

\newpage

\bibliography{swimming}
\end{document}